\documentclass[a4paper,11pt]{article}
\usepackage{jheppub}
\usepackage[T1]{fontenc}
\usepackage{graphicx}
\usepackage{svg}
\usepackage{slashed}
\usepackage{bbm}
\usepackage{changes}
\definechangesauthor[name={WQ Chen}, color=red]{wqc}

\title{Fate of Quantum Anomalies for 1d lattice chiral fermion with a simple non-Hermitian Hamiltonian}

\author[a,b]{Wei-Qiang Chen,}
\author[c]{Yong-Shi Wu,}
\author[a,b,d]{Wenjie Xi,}
\author[a,b]{Wei-Zhu Yi,}
\author[a,e]{Gen Yue}

\affiliation[a]{Department of Physics and Shenzhen Institute for Quantum Science and Engineering, Southern University of Science and Technology, Shenzhen 518055, China}
\affiliation[b]{Shenzhen Key Laboratory of Advanced Quantum Functional Materials and Devices, Southern University of Science and Technology, Shenzhen 518055, China}
\affiliation[c]{Department of Physics and Astronomy, University of Utah, Salt Lake City, UT 84112, USA}

\affiliation[d]{Department of Physics and HKU-UCAS Joint
Institute for Theoretical and Computational Physics,
The University of Hong Kong, Pokfulam Road, Hong Kong, China}
\affiliation[e]{Department of Physics, The Chinese University of Hong Kong, Sha Tin, New Territories, Hong Kong, China}

\abstract{

It is generally believed that the 1+1D model for a single chiral fermion does not exist by itself alone on lattice. The obstruction to such a lattice realization is the failure to reproduce the quantum anomalies of a chiral fermion in continuum.
The conventional way to escape is to associate the anomalous 1d system with a 2d bulk, which is in a topologically non-trivial state, as the boundary of the latter. 
In this paper, we propose a 1+1D chiral fermion model on 1d spatial lattice, {\it standing alone} -- without being associated with a 2d bulk -- with a simple {\it non-Hermitian} hopping Hamiltonian. We demonstrate, using various methods, that the model possesses the same chiral anomaly and gravitational anomaly as in continuum theory. Furthermore, with appropriate parameters, the low energy effective theory of the model remains a field theory for unitary chiral fermions. The essential reason for the success is that the usual "doubled" fermion mode with opposite chirality is rapidly damped out because of non-Hermicity of the Hamiltonian.      
}

\begin{document}

\maketitle

\section{Introduction}

Lattice models play a ubiquitous role, as a discrete formulation of quantum field theory, in particle physics as well as in condensed matter physics. In particle physics, the lattice is considered as the ``most physical'' way of regularizing a continuum quantum field theory. In condensed matter physics, especially for crystalline systems, the lattice models are viewed as being more close to the "microscopic" reality. However, the relationship between the lattice and continuum formulations of a quantum field theory can be very subtle and complex, particularly when dealing with the so-called topological properties in the two formulations.\footnote{Readers will see more below.}

In mathematics, topological properties are defined as the ones invariant under continuous deformations. However, it does not doom to be impossible to define a topological property in a discrete setting. A classical example is the Euler characteristic defined on a triangulation of a closed smooth surface. On the other hand, indeed in the literature there is a well-known No-go theorem for realizing the chiral fermion on a lattice: In accordance with the famous Nielsen-Ninomiya theorem \cite{Nielsen1981}, a chiral fermion theory in 1+1 dimensions does {\it not} exist {\it by itself alone} on the square lattice, as long as it has an Hermitian Hamiltonian and respects chiral symmetry and locality. 

In fact, it has been realized that the reason underlying this No-go Theorem is closely related to the so-called quantum anomalies, including the chiral anomaly \cite{Adler1969,Bell1969} and the gravitational anomaly \cite{Witten1984}, for chiral fermions. In the proof of the No-go Theorem\cite{Nielsen1981}, it is shown that accompanying each chiral fermion mode, there is always another (the so-called "doubled") fermion mode with opposite chirality emerging naturally in the lattice model, which always cancels the quantum anomalies of the former. This is known as the famous "fermion doubling" phenomenon, which is the essence of the Nelson-Ninomiya theorem. To avoid the fermion doubling problem, the conventional way is to realize a quantum field theory on d-dimensional spatial lattice with quantum anomaly as the boundary of a (d+1)-spatial-dimensional bulk lattice model. For example, a 1+1D chiral fermion\footnote{In our convention, d stands for spatial dimension, while D for spacetime dimension.} can be realized as the boundary of a 2d integer quantum Hall (IQH) system\cite{Stone1991}, where the bulk Hall conductance gives rise to the anomaly flow from bulk to the boundary, so that charge conservation is respected for the bulk plus boundary combined system. 

There have been also many other proposals to realize chiral fermions on a lattice as a standing-lone model, not as the boundary of certain bulk theory. But this normally requires some additional complicated circumstances: Either the chiral symmetry is broken\cite{wilson1976erice,PhysRevD.16.3031}, or it is realized with a non-local Hamiltonians \cite{PhysRevD.14.487, demarco2018single}. Alternatively, one may add some kinds of interactions \cite{Eichten:1985ft,PhysRevD.99.111501,10.1093/ptep/ptz055,wang2013non,demarco2017novel,Zeng_2022,Wang_2022}, or requires some additional nontrivial bulk\cite{Kaplan_1992,demarco2022chiral}, and so on. 

In this paper we will undertake the task to develop a new formulation of the {\it standing-alone} (1+1)D lattice model for a chiral fermion, without a 2d bulk as well as without any above-mentioned extra complications. Our idea is to explore the fate of quantum anomalies in the 1d lattice model for a chiral fermion with a {\it non-Hermitian} Hamiltonian, which violates the Hermiticity -- a key prerequisite of the Nelson-Ninomiya No-go Theorem. In recent years, the study of non-Hermitian systems has attracted a lot of attention in the community of theoretical physics; see, for example, the references  \cite{PhysRevLett.123.206404,PhysRevLett.116.133903,PhysRevLett.118.040401,PhysRevLett.121.026808,PhysRevLett.121.086803,PhysRevLett.123.066405,PhysRevLett.124.046401,PhysRevLett.124.056802,PhysRevLett.124.066602,PhysRevLett.124.250402,PhysRevX.4.041001,PhysRevX.9.041015,demarco2018single}. Our idea here is inspired by an observation made by Nagata and one of the authors in ref. \cite{NagataWu2008} that the usual U(1) Chern-Simons gauge theory can be reformulated on a specific 3D lattice with a non-Hermitian Hamiltonian.\footnote{Finally, in that paper, adding the Hermitian conjugate results in a lattice model for {\it doubled} Chern-Simons theory.}

More concretely, we consider the simplest lattice model for 1d chiral fermions, that hop only to the left nearest neighboring site. Namely we discretize the spatial derivative of the fermion field as follows:   
\begin{equation}
\label{oriented}
\Psi_L^{\dagger}\partial_x \Psi_L \rightarrow \frac{1}{a} c^{\dagger}_{L,j}\lbrack c_{L,j+1}- c_{L,j}\rbrack. 
\end{equation}
so that the corresponding hopping Hamiltonian is non-Hertimian. We will show that 
with parameters in some appropriate range, this non-Hermitian lattice model realizes  a continuum theory with quantum anomaly in same spatial d-dimension. 

We organize our paper as follows. At the beginning, we briefly review the chiral anomaly and gravitational anomaly in the theory of 1+1-dimensional chiral fermion. Then we construct a non-Hermitian one-dimensional lattice model for the left-moving chiral fermion.  We will demonstrate that our model possesses the two anomalies from several aspects. Furthermore, a field theory of chiral fermion which is stable in a large time scale emerges in continuum limit and low energy regime of the model.
At last, we check that the chiral fermion is stable against perturbations.
Our model provides an approach to realize quantum anomaly in lattice model in the same spatial dimension with the help of non-Hermiticity. Our study may also provide an alternative possibility, besides the spontaneous symmetry breaking, for the origin of some phenomena with nonzero chirality in nature.

\section{Review of Quantum Anomalies in 1+1D Chiral Fermion}
The action for a 1+1 dimensional chiral fermion field theory in Minkowski space is given by \begin{equation}
    \label{contiAction}
    S = \int dt dx\  i\bar{\Psi}(x,t)(\gamma^\mu(\partial_\mu+iA_\mu))\Psi(x,t) 
\end{equation}
where we take the fermion velocity $v_F=1$, $\gamma^1  = -i\sigma_2$, $\gamma^0 =  \sigma_1$, $\gamma_5 = \gamma^0\gamma^1 = \sigma_3,\mu=0,1, \bar{\Psi}=\Psi^\dagger\gamma^0$,  .
A 1+1D left(right) moving chiral fermion field $\Psi_{L(R)}(x,t)$ can be achieved by projecting the Dirac spinor $\Psi(x,t) = (\Psi_R(x,t),\Psi_L(x,t))^{T}$ with the projection operator $P_{L,R}=(1\mp \gamma_5)/2$.
The chirality is given by the eigenvalue of $\gamma_5$, which is -1 for $\Psi_{L}(x,t)$ and +1 for $\Psi_{R}(x,t)$.

\subsection{Chiral Anomaly}

\label{sec:chiral_anomaly}
We consider the theory in Euclidean space. The Euclidean action is
\begin{equation}
    S_E = \int d\tau dx\ i\bar{\Psi}(x,\tau)(\gamma^\mu(\partial_\mu+iA_\mu))\Psi(x,\tau), 
\end{equation}
where $\gamma^2 = i\gamma^0 =  i\sigma_1$,$A_2 = -iA_0$, $x^2 = \tau = it, \mu = 1,2, \bar{\Psi}=\Psi^\dagger$.
It is invariant under the infinitesimal chiral transformation:
\begin{equation}
\label{ChiralTransformation}
    \Psi(x,\tau)\rightarrow \exp(i\alpha \gamma_5)\Psi(x,\tau),
    \bar{\Psi}(x,\tau)\rightarrow\bar{\Psi}(x,\tau)\exp(i\alpha \gamma_5).
\end{equation}
By Noether's theorem, this leads to a conserved axial current $j^\mu_5(x,\tau)= \bar{\Psi}(x,\tau)\gamma^\mu\gamma_5\Psi(x,\tau)$ and $    \partial_\mu j_5^\mu = 0
$. 

However, the axial current is not really conserved at the quantum level because the partition function is not invariant under chiral transformations; this phenomenon is called chiral anomaly. The chiral anomaly can be observed by making a local chiral transformation, where the parameter $\alpha$ in \eqref{ChiralTransformation} becomes $\alpha(x, \tau)$.  Then the path integral measure $
    d\mu = \Pi_x\mathcal{D}[A_\mu(x)]\mathcal{D}[\Psi(x)]\mathcal{D}[\bar{\Psi}(x)]
$
transforms as:
\begin{equation}
\label{ChiralAnomaly}
    d\mu \rightarrow d\mu\exp[-2i \int d\tau dx\ \alpha(x,\tau)\sum_n\varphi_n^\dagger\gamma_5\varphi_n],
\end{equation}
where $\varphi_n$s are the orthonormal eigenvectors of the Euclidean Dirac operator $\tilde{\slashed{D}}$ with eigenvalue $\lambda_n$. 
With the Ward-Takahashi identity\cite{peskin2018introduction},
 we have \begin{equation}
     \label{AxialCurrent}
     \partial_\mu j^\mu_5 = 2i\sum_n\varphi_n^\dagger\gamma_5\varphi_n.
 \end{equation}

The chiral anomaly is actually related to the Atiya-Singer index theorem
\cite{bertlmann2000anomalies}:
\begin{align}
\label{A-S index}
\mathrm{index}(\tilde{\slashed{D}}_R)= n_+-n_- =  \int dx d\tau\sum_n \varphi^\dagger_n  \gamma_5  \varphi_n = -\frac{1}{2\pi} \int dx d\tau F,
\end{align}
where $\mathrm{index}(\tilde{\slashed{D}}_R)$ is the index of $\tilde{\slashed{D}}_R$. $n_{\pm}$ are the numbers of zero modes of $\tilde{\slashed{D}}$ with chirality $\chi=\pm 1$ respectively.  $F$ is the field strength of the $U(1)$ gauge field.
Eqn. \eqref{A-S index} can be proved as follows.  The first equal sign is the definition of the index of $\tilde{\slashed{D}}_R$.  For the second equal sign, since $\{\tilde{\slashed{D}},\gamma_5\}=0$, we have  
\begin{equation}
\lambda_n  \varphi_{n}^\dagger  \gamma_5  \varphi_{n} =  \varphi_{n}  \gamma_5 \tilde{\slashed{D}}  \varphi_{n} 
=- \varphi_{n}^\dagger \tilde{\slashed{D}}  \gamma_5  \varphi_{n} =-\lambda_n  \varphi_{n}^\dagger  \gamma_5  \varphi_{n}.
\end{equation}
Thus, only the zero modes $\varphi_n^0$ of $\tilde{\slashed{D}}$ contribute to the sum.  The zero modes with chirality $\pm1$ are denoted as $\varphi_{n\pm}^0$ respectively, i.e.
\begin{equation}
    \gamma_5\varphi_{n\pm}^0=\pm \varphi_{n\pm}^0,
\end{equation}
and hence
\begin{equation}
\label{relation}
    \int dx d\tau\sum_n \varphi^\dagger_n  \gamma_5  \varphi_n = \int dx d\tau\sum_n {\varphi_{n+}^0}^\dagger \varphi_{n+}^0-\int dx d\tau\sum_n {\varphi_{n-}^0}^\dagger \varphi_{n-}^0=n_+-n_-
\end{equation}
Then by using Fujikawa regularization\cite{Fujikawa1979}, one can get
\begin{equation}
    \int d\tau dx \sum_n\varphi_n^\dagger\gamma_5\varphi_n = -\frac{1}{2\pi}\int _{\mathbb{R}^2} F.
\end{equation}
The Fujikawa's regularization procedure is to insert a factor  $\exp{(-\frac{\slashed{D}^2}{M^2})}$ to get a regularized sum $\sum_n \varphi_n^\dagger\gamma_5\exp{({-\frac{\slashed{D}^2}{M^2}})}\varphi_n$, $M\rightarrow \infty$. It can be evaluate by doing Fourier transformation $\tilde{\varphi}_n(k)$ of $\varphi_n(x,t)$. The $U(1)$ field strength term $F$ comes from the commutator $[D_\mu, D_\nu]=-iF_{\mu\nu}$, which is in the expansion of $\slashed{D}^2$. Finally, by doing the rescaling $k\rightarrow Mk$, one can get: \begin{equation}
    \sum_n\varphi_n^\dagger\gamma_5\varphi_n = M^2\int \frac{d^2k}{(2\pi)^2}e^{-k^\mu k_\mu}\text{Tr}(\gamma_5\exp{(\frac{i\gamma^\mu\gamma^\nu F_{\mu\nu}}{2M^2})}) = -\frac{1}{4\pi}\epsilon^{\mu\nu}F_{\mu\nu}
\end{equation}
Please note that according to the index theorem, $\mathrm{index}(\tilde{\slashed{D}}_R)$ is a topological number, i.e. it does not change as long as the gauge field $F$ is in the same topological equivalence class.

The non-conservation equation \eqref{AxialCurrent} of the axial current is just the local version of the index theorem: by integrating both sides, with 
the index theorem, we have

\begin{equation}
\label{AxialChargeAndIndex}
    \int d^2x \partial_\mu j_5^\mu =2(n_+-n_-) =2\ \mathrm{index}(\tilde{\slashed{D}}_R)=-\frac{1}{\pi}\int _{\mathbb{R}^2} F.
\end{equation}
Thus, the $\mathrm{index}(\tilde{\slashed{D}}_R)$ is an indicator of the chiral anomaly.

Another widely used way to verify the chiral anomaly in Minkowski space
is to put the system on a one-dimensional ring, and to study the spectral flow of the system under an adiabatic increase in the gauge flux, that threads the ring, from $0$ to $2\pi$. 
In Minkowski space, the conserved charge of chiral symmetry is 
\begin{equation}
\label{axial charge}
    Q_5 = \int dx j^0_5  = \int dx (\Psi_R^\dagger\Psi_R-\Psi_L^\dagger\Psi_L)=(Q_R-Q_L)
\end{equation}
On one hand, the axial charge should be conserved during the process because of the chiral symmetry.  On the other hand, as suggested by Eqn. \eqref{AxialChargeAndIndex}, the axial charge of a chiral theory will change by the chirality of the theory after the insertion of the flux.  This discrepancy is just the chiral anomaly.  And, the change of the axial charge can be demonstrated by checking how many states cross the zero energy during the adiabatic process, i.e. the spectral flow.

\subsection{Gravitational Anomaly}
A gravitational anomaly  is a phenomenon that the partition function is non-invariant under diffeomorphism transformation.
According to the conformal field theory (CFT) \cite{blumenhagen2009introduction}, the partition function of a complex chiral fermion (two Majorana fermions) on a 2D torus is given by 
\begin{equation}
    Z(\tau,\bar{\tau})= tr(q^{L_0-\frac{c}{24}}\bar{q}^{\bar{L}_0-\frac{\bar{c}}{24}})
\end{equation}
where $L_0$($\bar{L}_0$) is a generator of the Virasoro algebra of the holomorphic (antiholomorphic) part; $c$ and $\bar{c}$ are the central charge of the holomorphic and antiholomorphic part, respectively.  We stipulate that the left-moving free fermion has $\bar{c}=1$, while the right-moving free fermion has $c=1$. $q=e^{2\pi i\tau}$ is a parameter, where $\tau$ is a complex number characterizing the torus.  Please note that those $\tau$'s differing by a modular transformation
\begin{equation}
    \tau \sim \frac{a\tau+b}{c\tau+d}, \text{\quad with} 
    \begin{pmatrix}
     a&\ b\\c&\ d
\end{pmatrix} \in SL(2,\mathbb{Z})
\end{equation}
describe the same torus, and hence the partition function should be invariant under the modular transformations.  Under a special modular transformation $T: \tau\rightarrow \tau +1$, the partition function becomes
    \begin{equation}
    \label{GA indicator}
    Z\longrightarrow Z\exp\left(2\pi i\frac{c-\bar{c}}{24}\right).
\end{equation}
Since the chiral central charge of the left-moving complex fermion theory is $c_-:=c-\bar{c}=-1$, its partition function is non-invariant under the $T$ transformation.  And we say that the theory has a gravitational anomaly.

\section{The Lattice Model and Its Spectrum}

\begin{figure}
\centering
\includegraphics[width=8.5cm]{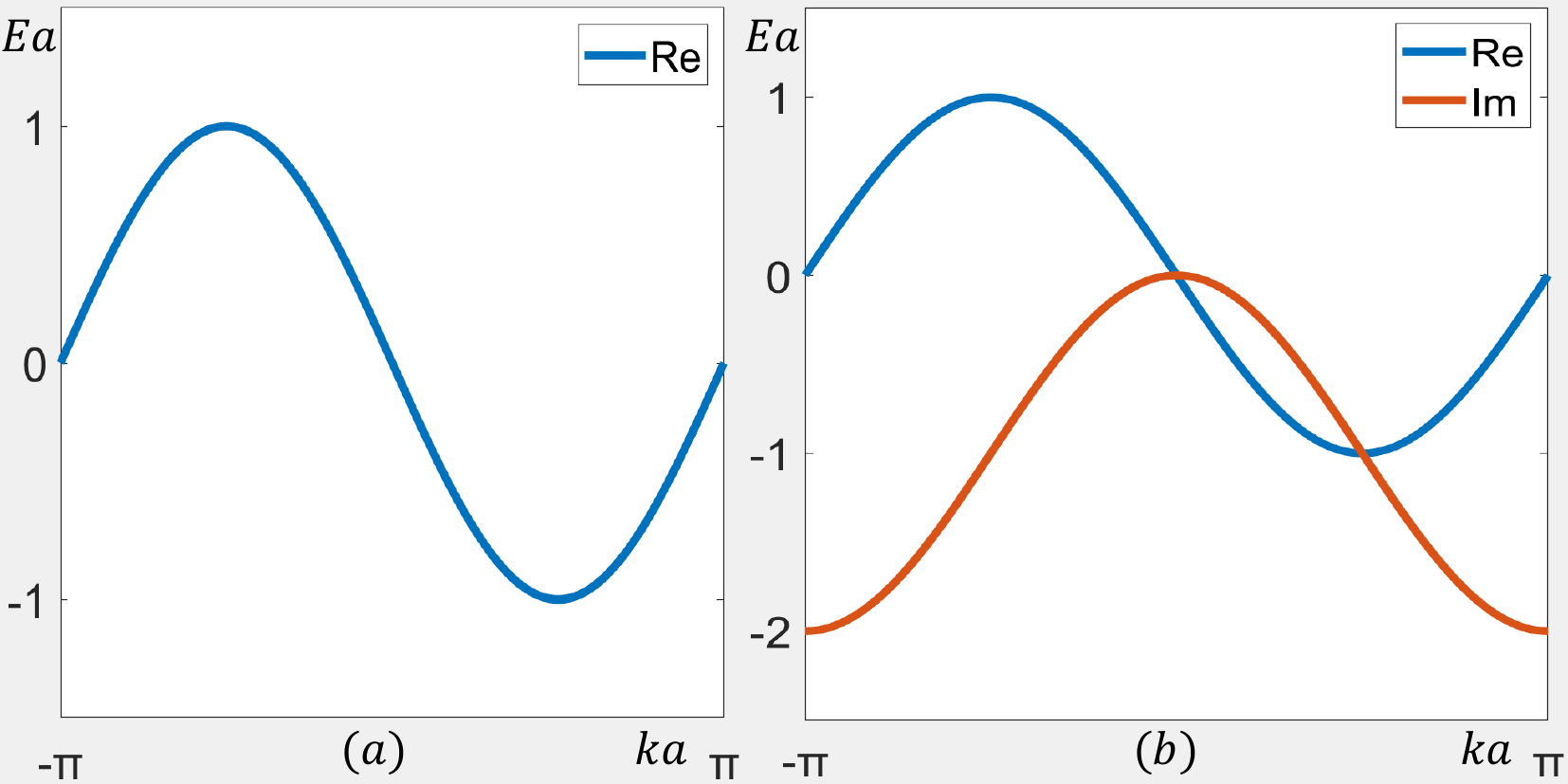}
\caption{\label{Spectrum}
The energy dispersion of $(a)$ the symmetric and $(b)$ the oriented lattice model, respectively, with periodic boundary conditions and $A_1=0$. Here the blue and red line are, respectively, the real and imaginary part of the energy.}
\end{figure}

Now we construct a lattice model for a left-moving fermion in the Hamiltonian approach. We assume the 1d spatial dimension is discretized into a set of $N$ points on a circle with lattice constant $a$, while the time remains continuous. Recall that the continuum Hamiltonian of the left-moving fermion to be discretized is given by 
\begin{align}
\label{Ham_conti}
H_L=\int dx\ {\Psi}^\dagger_L(x,t) \lbrack i\partial_x - A_1 \rbrack \Psi_L (x,t),
\end{align}
where, for later convenience of computing chiral anomaly, we have introduced the gauge potentials: $A_0 = 0, A_1 = A_1(t)$. 
The corresponding lattice Hamiltonian is constructed by the following steps:
\begin{enumerate}
    \item Replace the field operator $\Psi_{L(R)}(x,t)$, $\Psi_{L(R)}^\dagger(x,t)$ in the Heisenberg's picture by annihilation and creation operators of fermions $c_{L(R),j}$, $c_{L(R),j}^\dagger$ in Schr\"{o}dinger's picture, 
i.e
\begin{align}
    \Psi_{L(R)}(x,t) &\rightarrow c_{L(R),j} & \Psi^\dagger_{L(R)}(x,t) &\rightarrow c^\dagger_{L(R),j},
\end{align}
where $j$ is the lattice site index. We will impose periodic boundary conditions in the following discussions
\begin{align}
    \label{PBC}
        c_{L(R),N+j} = c_{L(R),j} & &        c_{L(R),N+j}^\dagger = c_{L(R),j}^\dagger\ ;
\end{align}  
Main results of this paper are qualitatively the same for anti-periodic boundary conditions. 
\item The gauge field is introduced by a phase factor in the hopping terms
\begin{equation}
    \label{GaugeFactor}
    e^{iaA_1} c_{L(R),j}^\dagger c_{L(R),j+1}.
\end{equation}
\item Replace the differential operator $\partial_1$ with some choice of the difference operator on the lattice. Different choices give rise to different lattice models. In general, a difference operator can be expressed as
\begin{align}
     \frac{1}{a}[\lambda ( c_{L,j+1}- c_{L,j}) + (1- \lambda) ( c_{L,j}- c_{L,j-1})].
\end{align}
\end{enumerate}

The discretization in eqn. \eqref{oriented} corresponds to the case with $\lambda=1$. 
 As mentioned in the introduction, such a choice can avoid the ``fermion doubling'' problem at the cost of a non-Hermitian Hamiltonian.  To demonstrate it, we will give a brief review on how the fermion doubling problem arises in the conventional approach of discretizing it to a lattice model.

In conventional approach, one select the symmetric choice $\lambda = 1/2$, which gives
\begin{align}
\label{eq:symmetric}
\Psi_L^\dagger\partial_x \Psi_L \rightarrow \frac{1}{2a}c_{L,j}^\dagger\lbrack c_{L,j+1}- c_{L,j-1}\rbrack,
\end{align}
and leads to a Hermitian lattice Hamiltonian
\begin{equation}
    \label{SymmetricHamiltonian}
    H_{\text{sym}} = \frac{i}{2a}(\sum_{j=1}^{N-1}e^{iaA_1}c_{L,j}^\dagger c_{L,j+1}-\sum_{j=2}^N e^{-iaA_1}c^\dagger_{L,j}c_{L,j-1}+ e^{iaA_1}c_{L,N}^\dagger c_{L,1}-e^{-iaA_1}c^\dagger_{L,1}c_{L,N}).
\end{equation}
In the following, we refer to this difference operator and the corresponding model as the symmetric difference operator and symmetric lattice model.
If $A_1$ is regarded as a time-independent constant, the spectrum can be expressed as
\begin{equation}
    \label{SymmetricSpectrum}
    E_{k}^{\text{sym}} = -\frac{\sin((k+A_1)a)}{a},
\end{equation}
where $k=\frac{2\pi n}{Na}$, $n\in \mathbb{Z}$; the superscript `sym' stands for `symmetric'.

However, this lattice model has the fermion doubling problem, i.e. besides a branch of left-moving fermion near $k \sim 0$, there is a second (or doubled) fermion near $k \sim \pi/a$ as shown in fig.~\ref{Spectrum}(a).  Since  $c_{L,j}=\frac{(1-\gamma_5)}{2}\Psi_j$, the corresponding chirality $\chi$ must be -1. However, the energy of the doubled fermion is proportional to the momentum, which suggests that it is ``right-moving''. Therefore, the spectrum of the Hamiltonian is not chiral, i.e. it consists of two branches of both left-moving and doubled fermions, not merely the left-moving ones as one would have naively expected from the continuum Hamiltonian  \eqref{Ham_conti}. 
If we take the continuum limit $a\to 0, N\to \infty$ with $Na=L$ finite, it is known that the emergent doubled right-moving fermion has a contribution to the chiral anomaly, which exactly cancels that of the original left-moving fermion. Thus, the non-vanishing chiral anomaly in the continuum model \eqref{Ham_conti} is {\it not reproduced} in the discretized model defined by the symmetric Hamiltonian
\eqref{SymmetricHamiltonian}. This is the essence of the Nielsen-Ninomiya theorem.           
One may escape from the Nielsen-Ninomiya theorem by giving up the Hermiticity condition.\cite{KARSTEN1981,Chernodub2017}.  This can be understood with a proof of the Nielsen-Ninomiya theorem based on the Poincaré-Hopf theorem \cite{KARSTEN1981}. According to this theorem, the sum of indices of all the isolated zero modes in a 1D lattice model of chiral fermion is 0.  For a local, Hermitian, translation invariant model, the index of an isolated zero mode is defined by $1,-1$ for left-moving fermions and right-moving fermions, respectively.  Thus one must have equal numbers of left-moving and right-moving fermions to guarantee the zero of the sum of indices.  However, for non-Hermitian systems, the index of an isolated zero mode is 0 and is not related to zero mode's chirality. Therefore, it is possible to have only one left-moving fermions in a non-Hermitian lattice model.

Since the $\mathcal{PT}$- symmetry of a system guarantees a real spectrum and we need a complex spectrum to escape from the Nielsen-Ninomiya theorem, we must explicitly break the $\mathcal{PT}$- symmetry to obtain a Hamiltonian which describes a chiral fermion. In our case, the parity and time reversal transformations are defined, respectively, by 
\begin{align}
  \mathcal{P}:& c_{L, j} \rightarrow c_{L, N+1-j}, & c_{L, j}^\dagger &\rightarrow c_{L, N+1-j}^{\dagger},   \nonumber\\
  \mathcal{T}:& c_{L, j} \rightarrow c_{L, j}, & c_{L, j}^\dagger &\rightarrow c_{L, j}^{\dagger}, & i\rightarrow -i.
\end{align}
One can check the symmetric Hamiltonian \eqref{SymmetricHamiltonian} is $\mathcal{PT}$- symmetric.

When $\lambda=1$, \eqref{oriented} 
leads to a non-Hermitian lattice model
\begin{align}
\label{Hamiltonian}
{H}_{L}=\sum_{j=1}^{N-1}\frac{i}{a} e^{i a A_1} c_{L,j}^\dagger c_{L,j+1} - \sum_{j=1}^N \frac{i}{a} c_{L,j}^\dagger c_{L,j} +\frac{i}{a}e^{iaA_1}c_{L,N}^\dagger c_{L,1}.
\end{align}
Since this Hamiltonian breaks the $\mathcal{PT}$- symmetry explicitly, it is not surprising to find a complex spectrum:
\begin{equation}
\label{spectrumL}
    E_k^L  = \frac{i}{a} \left( e^{i a (k+A_1)}-1 \right).
\end{equation}
This model will be referred to as the oriented lattice model below, and we will show that it does describe a free left-moving fermion under appropriate conditions.

The model of a free right-moving fermion can be achieved by performing the parity transformation on the model \eqref{Hamiltonian},
\begin{equation}
\label{HamiltonianR}
    {H}_{R}=\sum_{j=2}^N\frac{i}{a} e^{-i a A_1} c_{R,j}^\dagger c_{R,j-1} - \sum_{j=1}^N \frac{i}{a} c_{R,j}^\dagger c_{R,j}+\frac{i}{a}e^{-iaA_1}c_{R,1}^\dagger c_{R,N},
\end{equation}
where we have replaced the subscript $L$ with $R$ to indicate that it is for a right-moving fermion. The spectrum of this model is also complex
\begin{equation}
\label{SpectrumR}
     E_k^R  = \frac{i}{a} \left( e^{-i a (k+A_1)}-1 \right),
\end{equation}
where $k=\frac{2\pi n}{Na}$, $n\in \mathbb{Z}$. The real part of its spectrum is opposite to \eqref{spectrumL}, while the imaginary part of these two Hamiltonians are the same. 
The Hamiltonian \eqref{HamiltonianR} can also be derived from the right-moving fermion field theory by choosing a different difference operator with $\lambda=0$
\begin{align}
\label{orientedR}
\Psi_R^\dagger \partial_x \Psi_R \rightarrow \frac{1}{a}c_{R,j}^\dagger\lbrack c_{R,j}- c_{R,j-1}\rbrack.
\end{align}

For a non-Hermitian system, it's necessary to introduce the so-called bi-orthogonal basis\cite{Brody_2013}. In the usual hermitian cases, the orthogonality of eigenstates is guaranteed by the hermiticity of Hamiltonians.  However, the eigenstates of a non-Hermitian Hamiltonian are in general not orthogonal to each other.  Instead, for a general non-Hermitian operator $A$, one can use the so-called bi-orthogonal basis\cite{Brody_2013}, which consists of the left eigenvectors $\{\langle v_{n}'|
\}$ and right eigenvectors $\{|v_{n}\rangle\}$ satisfying the following equations, 
\begin{align}
\label{bi-orthogonal}
        A |v_n\rangle &= \lambda_n |v_{n}\rangle,
        & \langle {v_n}'| A  &= \lambda_n \langle v_{n}'|,
        &\langle v_{n}'| v_m\rangle = \delta_{n,m}.
\end{align}
The second equation in \eqref{bi-orthogonal} implies that the hermitian conjugate of left eigenvectors are the eigenvectors of $A^\dagger$ \begin{equation}
    A^\dagger |{v_n}'\rangle = \lambda_n^*|{v_n}'\rangle.
\end{equation}
The bi-orthogonal basis also plays an important role in our following discussions.

We will now focus on the Hamiltonian \eqref{Hamiltonian} and its spectrum.  A similar analysis can be easily applied to the right-moving Hamiltonian \eqref{HamiltonianR}.  For a fermionic theory, the propagator of a fermion with a complex energy $E_k$ is given by $G(k,\omega)=1/(\omega-\mathrm{Re}E_k-i\mathrm{Im}E_k)$.
This suggests that the real part of the energy $\mathrm{Re} E_k$ corresponds to the ordinary energy of the particle, while the negative imaginary part $\mathrm{Im} E_k$ can be regarded as the inverse of lifetime, or a loss rate of the particle due to the coupling with the environments. Alternatively, the imaginary energy $\mathrm{Im} E_k$ can also be understood by considering the time evolution of a right eigenstate $|\psi_{k}\rangle$
\begin{align}
\label{TimeDependentState}
	|\psi_k(T)\rangle=e^{-iHT}|\psi_k\rangle=e^{-i\text{Re}(E_k)T}e^{\text{Im}(E_k)T}|\psi_k\rangle.
\end{align}
The the probability of finding the particle in the system 
\begin{equation}
\label{Probability}
    \langle \psi_{k}'(T)|\psi_{k}(T)\rangle = e^{2\text{Im}(E_k)T},
\end{equation}
decays with time if the imaginary part of the energy is negative. Note that $\langle \psi_k'(T)|$ is a left eigenstate. $T_0 = -1/\text{Im}(E_k)$ can be understood as the lifetime of the fermion. Physically, such a loss happens because of the coupling between the system and its environment.

Fig. 1 shows the energy dispersion for (a) the symmetric and (b) the oriented models, respectively. The real part of energy, $\mathrm{Re} E_k$, behaves similarly in both cases. This means that our (orient) model remains to have a left-moving fermion at $k\sim 0$ and a (doubled) right-moving fermion at $k \sim \pi/a$. However, these two fermion modes have very different lifetimes.  For the left-moving fermion and doubled fermion, we have
\begin{align}
\label{SpectrumLimit}
   E^L_{0,k} &\sim -k(1+i ka /2), & 
    E^{L}_{\pi,k} &\sim k(1 -2i/ka),
\end{align}
where $E^L_0$ and $E^L_{\pi}$ stand for the energy of left-moving fermion and doubled fermion, respectively. Thus the lifetime of left-moving fermion is much larger than that of the doubled fermion provided $a$ is small enough.  And in the continuum limit $a \rightarrow 0$, the left-moving fermion has infinite lifetime, while the lifetime of the doubled fermion vanishes.  This indicates that the lattice Hamiltonian will reduce to the Hamiltonian \eqref{Ham_conti} for a chiral fermion in the continuum limit and low energy regime.  Moreover, there is an emergent $\mathcal{PT}$ symmetry in the continuum limit and low energy regime, which guarantees a real spectrum of the continuous theory as we discussed above.

\section{Chirality and Chiral Symmetry}

An ideal lattice realization of the chiral fermion is normally expected to reproduce the chiral anomaly in the continuum theory. In this section, we will examine the fate of chiral anomaly in our lattice model by two approaches, the spectral flow and the lattice version of the index theorem.

\begin{figure}
\centering
\includegraphics[width=8.5cm]{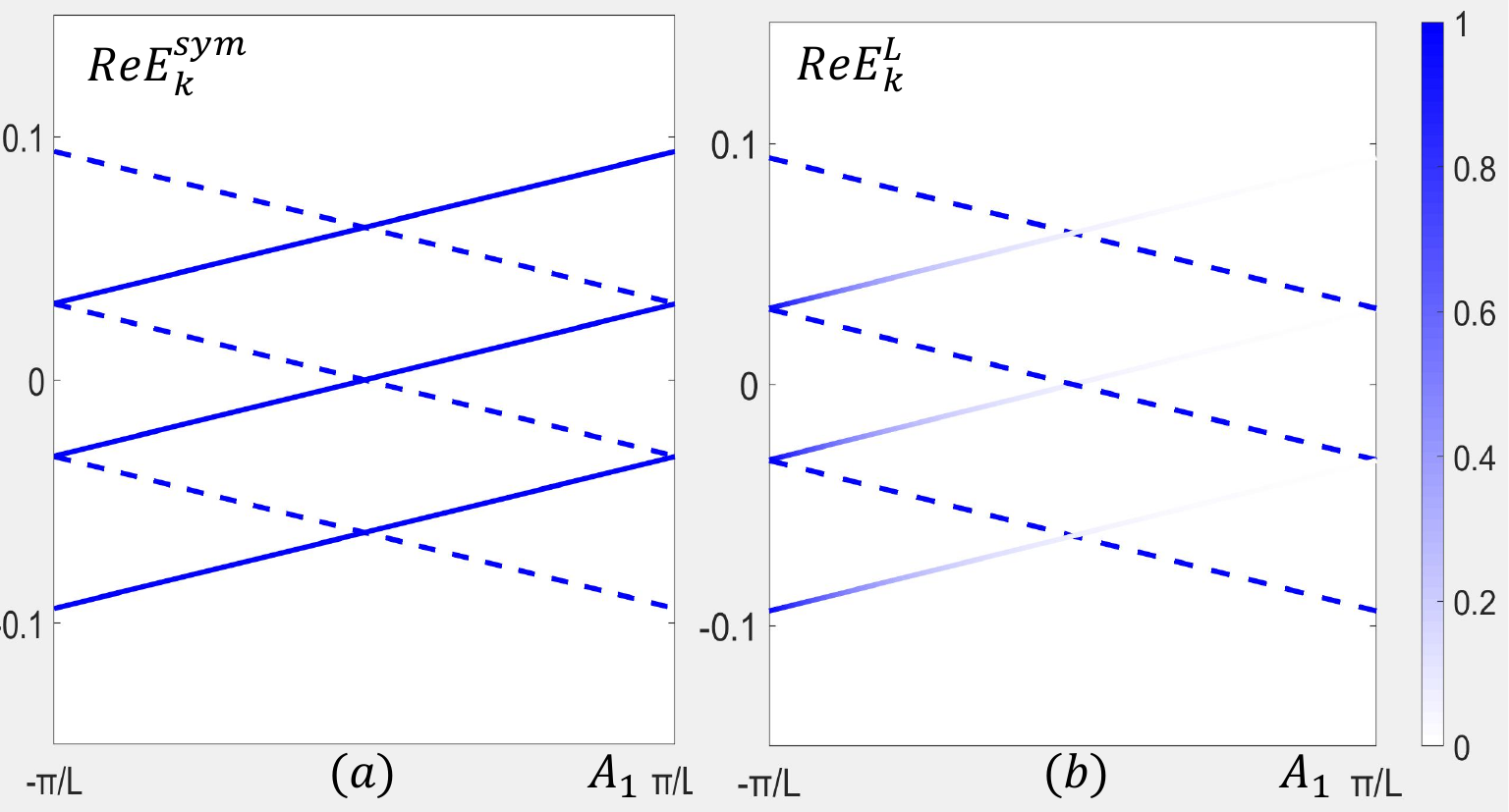}
\caption{\label{Fig_SF}
Illustration of spectral flow of the band driven by varying $A_1$ from $-\pi/L$ to $\pi/L$ adiabatically are shown in $(a)$ for the symmetric lattice model and $(b)$ for the oriented lattice model. Every branch is the energy of a certain eigenstate which varies with gauge field. Left-moving fermions are drawn in dash lines and doubled fermions are drawn in solid lines. Strength of the line implies the observation probability of these states. We set $a=1$, $N=100$.}
\end{figure}

\subsection{Spectral Flow}
\label{Spectral Flow}

 As discussed in Sec. \ref{sec:chiral_anomaly}, the spectral flow is a widely used approach to detect the chiral anomaly.
To observe the spectral flow, we slowly change $A_1$ from $-\pi/L$ at time $t = 0$ to $\pi/L$ at time $t=T$, and check how the energy of the fermions varies with $A_1$ (and hence with time).  Here we choose $L = Na$, where $N$ is total number of lattice sites.  $T$ should be large enough for an adiabatic process.  The resulting spectral flow for the symmetric and oriented lattice models is depicted in fig.\ref{Fig_SF} (a) and (b), respectively, where we choose $a= 1, N=100$.
Note that in the oriented lattice model case, the energy of fermion corresponds to the real part of the spectrum $\text{Re} E_k^L$ as we discussed in previous section, while the lifetime of the fermions, which is tracked by the imaginary part of the spectrum $\text{Im}E_k^L$, is presented with the strength of the lines.  For a given $A_1$, the energy of the fermions in the two models is the same and given by
\begin{equation*}
    E_{k}^{\text{sym}} = \text{Re}(E_k^L) = -\frac{\text{sin}((k+A_1)a)}{a},
\end{equation*} 
where $k = \frac{2\pi n}{Na}$. In the figures we depict only the evolution of the states with energy close to zero, i.e. three left-moving states (dash line) with $n=-1,0,1$ and three doubled fermion states (solid line) with $n=49,50,51$.

As we discussed in Sec. \ref{sec:chiral_anomaly}, physically, the change of $A_1$ corresponds to slowly inserting a $2\pi$ flux in the center of the ring.  The chiral anomaly requires that the total number of fermions should decrease by 1 after inserting the flux.  However, for the symmetric lattice model,
the number of doubled fermions increases by 1 and the number of left-moving fermions decreases by 1, hence the total number of fermions is invariant.  Therefore the symmetric lattice model does not have chiral anomaly.
For the oriented lattice model, there is an extra time-dependent decay factor $e^{2T \mathrm{Im} E}$. For $T$ is large enough
$ T \gg a$
 but not too large 
 $    T \ll \frac{{l_0}^2}{a}$,
where $l_0$ is some finite infrared cut-off length scale, left-moving fermions are almost unaffected but doubled fermions vanish rapidly with time evolution as illustrated in fig. \ref{Fig_SF} (b). In this case, the total number of fermions decreased by 1 after inserting the flux, which is consistent with the correct chiral anomaly \eqref{Ham_conti}. 
The infrared cut-off $l_0$ is chosen to be smaller than $L$ to guarantee that only the modes around $k = 0$ survive the time evolution.  And we also choose it to be scale with $L$ to make sure that it goes to infinity in the thermodynamic limit $L \rightarrow \infty$.
\subsection{Index Theorem}

To show that our lattice model does have the same chiral anomaly as in the continuum theory, we will investigate the index of the non-Hermitian lattice operator $\slashed{D}_R$ in this section. We will work in Euclidean spacetime. With a derivation similar to that of the lattice Hamiltonians \eqref{Hamiltonian} and \eqref{HamiltonianR}, the Euclidean action of the Dirac fermion on a 1D spatial lattice reads
\begin{equation}
    \label{action}
    S_E = \int d\tau i \bar{\Psi} \slashed{D} \Psi,
\end{equation} 
where $\Psi = (\Psi_R, \Psi_L)^T = ((c_{R,1},\cdots,c_{R,N}),(c_{L,1},\cdots,c_{L,N}))^T$, $\bar{\Psi} = \Psi^\dagger$,
and $c_{L(R),j}$ are fermion annihilation operators of fermions introduced in the previous section.   
The lattice version of the Dirac operator is given by
\begin{equation}
\label{DiracOperator}
    \slashed{D} := \gamma^1\begin{pmatrix}
         D_1^R&0\\
         0& D^L_1
    \end{pmatrix}+\gamma^2\partial_{\tau}=\begin{pmatrix}
         0& -D^L_1+i\partial_\tau\\
         D^R_1+i\partial_\tau&0
    \end{pmatrix},
\end{equation}
where the covariant derivative operators $D_1^{L/R}$ are $N\times N$ matrices with matrix elements
\begin{align}
     {D_1^{L}}_{i,j} &= \frac{e^{ia A_1} \delta_{i+1, j} - \delta_{i,j}}{a},   & {D^R_1}_{i,j} &= -\frac{e^{-iaA_1}\delta_{i-1,j}-\delta_{i,j}}{a},
\end{align}
where $i$ and $j$ are indices of lattice sites. 
In particular, ${D_1^L}_{N,1} = \frac{1}{a}e^{iaA_1}$ and 
${D_1^R}_{1,N} = -\frac{1}{a}e^{-iaA_1}$. Here, we also use the periodic boundary condition.

Please notice that $\slashed{D}$ is also a non-Hermitian operator. 
Thus, we introduce the bi-orthogonal basis $ \{\varphi_{n}\} $  and  $\{\varphi_{n}'\} $,
which satisfies \begin{equation}
\label{LRBasis}
\begin{split}
    &\slashed{D}\varphi_{n}(\tau) = \lambda_n \varphi_{n}(\tau)\\& \slashed{D}^\dagger \varphi_{n}'(\tau) = \lambda_n^* \varphi_{n}'(\tau)\\& \int d\tau\ {\varphi_{n}'}^\dagger(\tau)\varphi_{m}(\tau) = \delta_{n,m}.
\end{split}
\end{equation}

The action \eqref{action} respects chiral symmetry, as the Dirac operator is invariant under a lattice-version infinitesimal chiral transformation
\begin{equation}
\label{eq:chiral trans}
\Psi \rightarrow e^{i \alpha \gamma_5}\Psi, \ \bar{\Psi} \rightarrow \bar{\Psi} e^{i \alpha \gamma_5}.
\end{equation}
and the path integral measure change as \begin{equation}
    d\mu\rightarrow d\mu\exp{(-2i\alpha\int d\tau \sum_n {\varphi_{n}'}^\dagger\gamma_5\varphi_{n})},
\end{equation} 
the extra phase is the expected anomaly term.

We can define the corresponding Weyl operators as $\slashed{D}_{R,L} \equiv \slashed{D}(1\pm\gamma_5)/2$. With all above mentioned definitions, we are now going to prove the lattice-version index theorem

\begin{align}
\label{A-S index lattice}
\mathrm{index}(\slashed{D}_R)=n_+-n_-=\int d\tau \sum_n{\varphi_{n}'}^\dagger \gamma_5 \varphi_{n}=-\frac{1}{2\pi} \int_M F,
\end{align}
The first equal sign in equation \eqref{A-S index lattice} is valid for all elliptic operators. Since the generalized Weyl operator $\slashed{D}_R$ is elliptic, the first equal sign still holds in this case. The proof of the second equal sign is similar to that of \eqref{relation}.
By applying the normalization in eqn. \eqref{LRBasis}, we have
\begin{align}
\int d\tau \sum_n{\varphi_{n}'}^\dagger \gamma_5 \varphi_{n}=n_+-n_-
\end{align}

The third equal sign can be proved in a way similar to ref. \cite{Fujikawa1979}. Since $\int d\tau \sum_n{\varphi_{n}'}^\dagger \gamma_5 \varphi_{n}$ does not depend on the choice of basis, we choose to use the bi-orthogonal eigenstates of $D_1$, denoted by  $\{\phi_{n}'\},\{\phi_n\}$, to calculate the sum. Follow Fujikawa's approach \cite{Fujikawa1979},
\begin{equation}
\label{eqn:fujikawa}
\begin{split}
    \int d\tau \sum_n {\phi_{n}'}^\dagger(\tau)\gamma_5 \phi_{n}(\tau) &= \int d\tau \sum_j a \sum_n {\phi_{n,j}'}^\dagger(\tau)\gamma_5 \phi_{n,j}(\tau)\\
    & =  \lim_{M\rightarrow\infty}\int d\tau \sum_j a  \sum_n {\phi_{n,j}'}^\dagger(\tau) \gamma_5\exp{(-\frac{\slashed{D}^2}{M^2})}\phi_{n,j}(\tau),
\end{split}
\end{equation}
where
\begin{equation}
\label{CommutationRelation}
    \slashed{D}^2 = D_\mu D^\mu +\frac{1}{2}\gamma^\mu\gamma^\nu[D_\mu, D_\nu] = D_\mu D^\mu+\gamma^1\gamma^2[D_1, D_\tau].
\end{equation}

Since the spectrum of $D_1$ is just the energy of right and left-moving chiral fermions multiplied $\pm i$, the eigenstates $\{\phi_{n}\}$ can be divided into two sectors $\{\phi_{k}^R\}$ and $\{\phi_{k}^L\}$ which satisfy
\begin{equation}
\begin{split}
        &D_1\phi_{k}^R = iE_k^R\phi_{k}^R,\\
         &D_1\phi_{k}^L = -iE_k^L\phi_{k}^L,
\end{split}
\end{equation}
where $E_k^R$ and $E_k^L$ are given by \eqref{spectrumL} and \eqref{SpectrumR} respectively. 
And we have
\begin{equation}
\label{OrientedCM}
\begin{split}
     & [D_1, D_\tau]\phi_{k,j}^R = -(i\partial_\tau E^R_{k})\phi_{k,j}^R=-i\partial_\tau A_1 e^{-i(k+A_1)a}\phi_{k,j}^R,\\
     & [D_1, D_\tau]\phi_{k,j}^L = (i\partial_\tau E_{k}^L)\phi_{k,j}^L=-i\partial_\tau A_1 e^{i(k+A_1)a}\phi_{k,j}^L,
\end{split}
\end{equation}
where $j$ is the label of lattice sites. In the case with a tiny $a$, we may keep only the terms independent of $a$ and have
\begin{equation}
\label{CR}
[D_1, D_\tau] = -i\partial_\tau A_1.
\end{equation} 

Eqn. \eqref{eqn:fujikawa} be evaluated by introducing the Fourier transformation $\tilde{\phi}_{n,q}(\omega)$ of the eigenstate $\tilde{\phi}_{n,j}(\tau)$, where
\begin{equation}
    \phi_{n,j}(\tau) = \int \frac{d\omega}{2\pi} \frac{1}{L}\sum_q e^{iqja+i\omega\tau} \tilde{\phi}_{n,q}(\omega).
\end{equation}
$\tilde{\phi}_{n,q}(\omega)$ satisfies the following completeness condition, 
\begin{equation}
    \sum_n {{\tilde{\phi'     }}_{n,q'}}^\dagger (\omega')\Gamma \tilde{\phi}_{n,q}(\omega) = \text{Tr}(\Gamma)2\pi L\delta_{q,q'}\delta(\omega-\omega'),
\end{equation}
for any $2\times 2$ matirx $\Gamma$.  
Then we have
\begin{equation}
\begin{split}
     &\int d\tau \sum_n {\phi_{n}'}^\dagger(\tau)\gamma_5 \phi_{n}(\tau) \\
     &= \lim_{M\rightarrow \infty}\int d\tau \sum_j a\ \int \frac{d\omega}{2\pi}\frac{1}{L}\sum_q \text{Tr}(\gamma_5e^{-iqja-i\omega\tau}\exp{(-\frac{D_\mu D^\mu}{M^2}-\frac{\gamma^1\gamma^2[D_1,D_\tau]}{M^2})}e^{iqja+i\omega\tau}).
\end{split}
\end{equation}

Since $e^{-i\omega \tau}D_\tau e^{i\omega\tau} = D_\tau+i\omega, 
       e^{-iqja}D_1 e^{iqja} = D_1+ iq$,
we have
\begin{equation}
\label{NonHermitianHelp}
\begin{split}
 &\int d\tau \sum_n {\varphi_{n}'}^\dagger(\tau)\gamma_5 \varphi_{n}(\tau) = \int d\tau \sum_n{\phi_{n}'}^\dagger(\tau)\gamma_5 \phi_{n}(\tau)  \\
 & = \lim_{M\rightarrow \infty}\int d\tau \sum_j a \int \frac{d\omega}{2\pi}e^{-\frac{\omega^2}{M^2}}\frac{1}{L}\sum_q e^{-q^2/M^2}\text{Tr}(\gamma_5\exp{(-\frac{2iq_\mu D^\mu}{M^2}}-\frac{D_\mu D^\mu}{M^2}+\frac{i\gamma^1\gamma^2\partial_\tau A_1}{M^2}))\\
 & = \lim_{M\rightarrow \infty} M^2 \int d\tau \int dx \int \frac{d\omega}{2\pi}\int \frac{dq}{2\pi} e^{-q^2-\omega^2} \text{Tr}(\gamma_5\exp{(-\frac{2iq_\mu D^\mu}{M}}-\frac{D_\mu D^\mu}{M^2}+\frac{i\gamma^1\gamma^2\partial_\tau A_1}{M^2})),
\end{split}
\end{equation}
from the second line to the third line, we take the continuum limit and do a rescaling $q\rightarrow Mq, \omega \rightarrow M\omega$. Since
the Dirac matrices statisfy \begin{equation}
    \text{Tr}(\gamma_5)=0, \quad
   \text{Tr}(\gamma_5\gamma^1\gamma^2) = -2i,
\end{equation}
we have \begin{equation}
    \int d\tau \sum_n {\varphi_{n}'}^\dagger(\tau)\gamma_5 \varphi_{n}(\tau) = \int d\tau \int dx \frac{1}{4\pi} 2\partial_\tau A_1 = -\int d\tau dx \frac{1}{4\pi}\epsilon^{\mu\nu}F_{\mu\nu} = -\frac{1}{2\pi}\int F.
\end{equation}
Thus the index theorem \eqref{A-S index lattice} is correct in our non-Hermitian lattice model.

As discussed in Sec. \ref{sec:chiral_anomaly}, the lattice index theorem \eqref{A-S index lattice} means that our lattice model is able to reproduce the correct chiral anomaly in the continuum theory. On the other hand, since the symmetric lattice model has no chiral anomaly, the lattice index theorem \eqref{A-S index lattice} should be invalid for the symmetric difference operators.  In the following, we will provide a brief explanation on how the oriented lattice model and symmetric lattice model are different and how the non-Hermiticiy of the oriented lattice model helps.

By definition,  only the zero modes of $\slashed{D}$ contribute to $\mathrm{index}(\slashed{D}_R)$.  The eigenvalues $\lambda$ of $\slashed{D}$ satisfy
\begin{equation}
    \lambda^2 = (\omega-iE_k^L)(\omega-iE^R_k),
\end{equation}
where $E_k^L$ and $E_k^R$ are the energy spectrum of lattice left-moving fermion and right-moving fermion, respectively.

For the symmetric lattice model, the zero modes correspond to $\omega=0,E_k^L = E_k^R=0$.  For the left-moving fermion, both the states around $k=0$, which corresponds to the left-moving fermion in continuum limit, and the doubled femrion states around $k= \pi/a$ contribute to the index.  That is the reason why the index theorem \eqref{A-S index lattice} fails in the symmetric lattice case.

Things become different in the oriented lattice case, where $E_k^L$ and $E_k^R$ are given by eqn. \eqref{spectrumL} and \eqref{SpectrumR} respectively.
Thus the zero modes correspond to
$\text{Re}(E_k^L)=\text{Re}(E_k^R)=0$, $\omega=-\text{Im}(E_k^L)=-\text{Im  }(E_k^R)$. Again for the oriented lattice model of left-moving fermions, $\text{Re}(E_k^L)=0$ corresponds to the left-moving fermion around $k=0$ with $\omega = -\text{Im}(E_k^L)= -k^2a/2 \sim 0$, and the doubled fermion around $k=\frac{\pi}{a}$ with $\omega = -\frac{2}{a} \gg 1$.  However, the factor $e^{-\omega^2}$ in eqn.\eqref{NonHermitianHelp} strongly suppresses the contribution of the doubled fermion, and hence only the left-moving fermion at $k \sim 0$ contribute to the integral.  Thus, one can reproduce the valid chiral anomaly with the index theorem \eqref{A-S index lattice} being valid as well.

\section{The Gravitational Anomaly}

According to eqn.\eqref{GA indicator}, the chiral central charge of 1D chiral fermion is an indicator of the gravitational anomaly.  For our lattice model, if the lattice constant $a$ is small enough, we may consider only the states around $k = 0$ and $k = \pi/a$, whose spectrums are given by \eqref{SpectrumLimit}.  If we ignore the imaginary part of the energy, the spectrum is the same as the free complex fermion CFT, where the $E^{L}_{\pi}$ corresponds to the holomorphic mode with $c=1$ and $E^L_0$ corresponds to the anti-holomorphic mode with $\bar{c}=1$.  As we discussed before, the imaginary part of the energy corresponds to the lifetime of the fermions.  Thus in the continuum limit $a\to0$, the holomorphic mode is damped 
out rapidly, and only the anti-holomophic mode survives.  Therefore, it should correspond to a CFT with chiral central charge $ c-\bar{c} = -1$, which indicates that our non-Hermitian lattice model has the same gravitational anomaly as the 1D left-moving chiral fermion in the continuum.

\begin{figure}[h]  
	\centering
	{\includegraphics[width=3.8in,height=2.9
in]{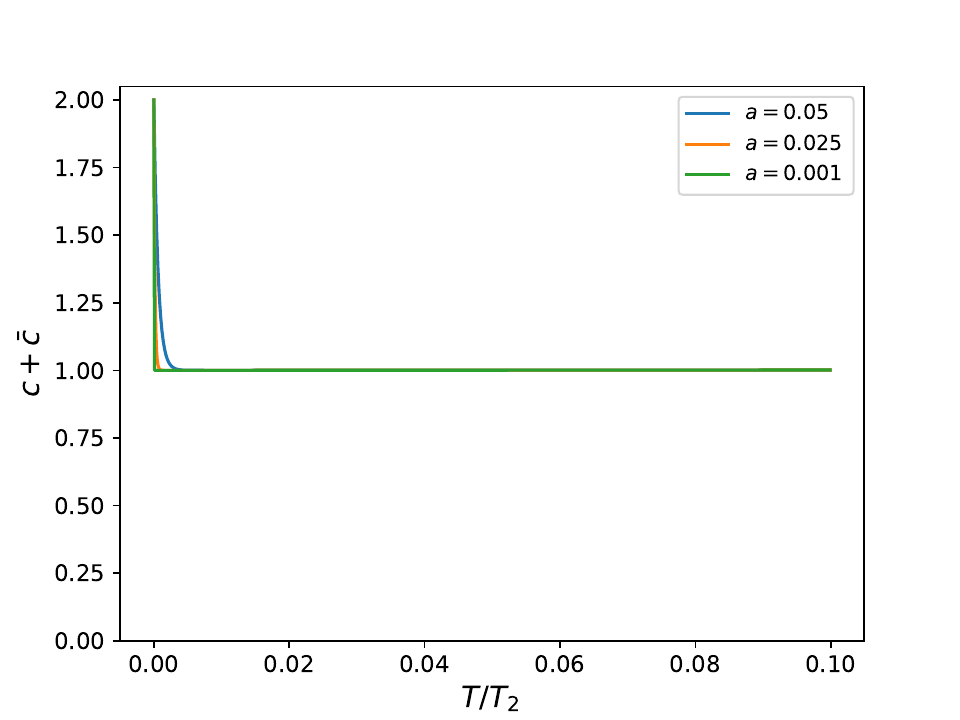}}
 	\caption{\label{TE} Illustrations of the time evolution of $c+\bar{c}$ in the interval $(0,\frac{1}{10}T_2)$ . $T_2$ is a characteristic time scale (see text for details) , $L=20$, 
 	$l_0=1$.}
\end{figure}

To make it clearer, we calculate numerically the difference of ground state energy, $\Delta E$,  between the case with periodic boundary condition(PBC) and the case with anti-periodic boundary condition (APBC). For a free fermion CFT, it is well-known that the vacuum energy difference between PBC and APBC is \cite{francesco2012conformal}
\begin{align}
\label{eq:cftdE}
\Delta E_{\text{vac}}=\frac{\pi}{4L}(c+\bar{c}),
\end{align}
where $L$ is the spatial length of the system and $c,\bar{c}$ count the number of doubled fermion branch and left-moving fermion branch. If the low energy effective theory of the oriented lattice model is a left-moving fermion, which corresponds to a fermion CFT with $\bar{c} = 1, c = 0$, we should have $\Delta E=\pi/4L$.
In our model, 
the time-dependent ground state energy for both PBC and APBC can be calculated by summing over the real part of the energy times the decaying probability factor given by \eqref{Probability} of all the occupied single particle states: 
\begin{align}
 \begin{split}
 \label{Et}
 	E(T)&=\sum_{k, occ} \text{Re}(E_k^L)\langle\psi_{k}'(T)|\psi_{k}(T)\rangle = \sum_{k, occ}e^{2\text{Im}(E_k^L)T}\text{Re}(E_k^L)\\&=-\sum_{k, occ}\frac{\sin ka}{a}e^{2\frac{\cos ka-1}{a} T},
 \end{split}
\end{align}
where $k=\frac{2\pi n }{L}$ for PBC, $k=\frac{2\pi(n+\frac{1}{2})}{L}$ for APBC, $L=Na$, $N$ is the number of sites. The subscript "occ" in the summation means that we only sum over all the occupied states.
By comparing with the result of CFT \eqref{eq:cftdE}, we can extract the the time evolution of $c+\bar{c}$, and the result is depicted in fig.\ref{TE}. 

Recall that in the continuum limit $a \rightarrow 0$, the lattice model starts from a non-chiral system and then rapidly evolves into a chiral system with only left-moving fermion. The chiral fermion state can last for a long time. 
Physically, there are two important time scales mentioned in \ref{Spectral Flow}.
First, at $T \gtrsim T_1=a$, the doubled fermions around $ka=\pi$ are damped out, only left-moving fermions survive. Note that $T_1$ tends to be $0$ in the continuum limit. While, at $T\approx T_2=\frac{{l_0}^2}{a}$, left-moving fermions around $ka=0$ begin to decay, which is the situation we hope to avoid. Fortunately, $T_2$ tends to infinity in the continuum limit. As a result, we should only consider the time region $T_1 \ll T\ll T_2$ when $a$ is finite. In this region, that the lattice model has same properties as a complex chiral fermion field theory with $c = 0, \bar{c} = 1$. Indeed, our numerical result $c + \bar{c} = 1$, as shown in fig. \ref{TE}, shows consistency with the field theory.

\section{Stability under Local Perturbations}

\begin{figure}
\centering
\includegraphics[width=9cm]{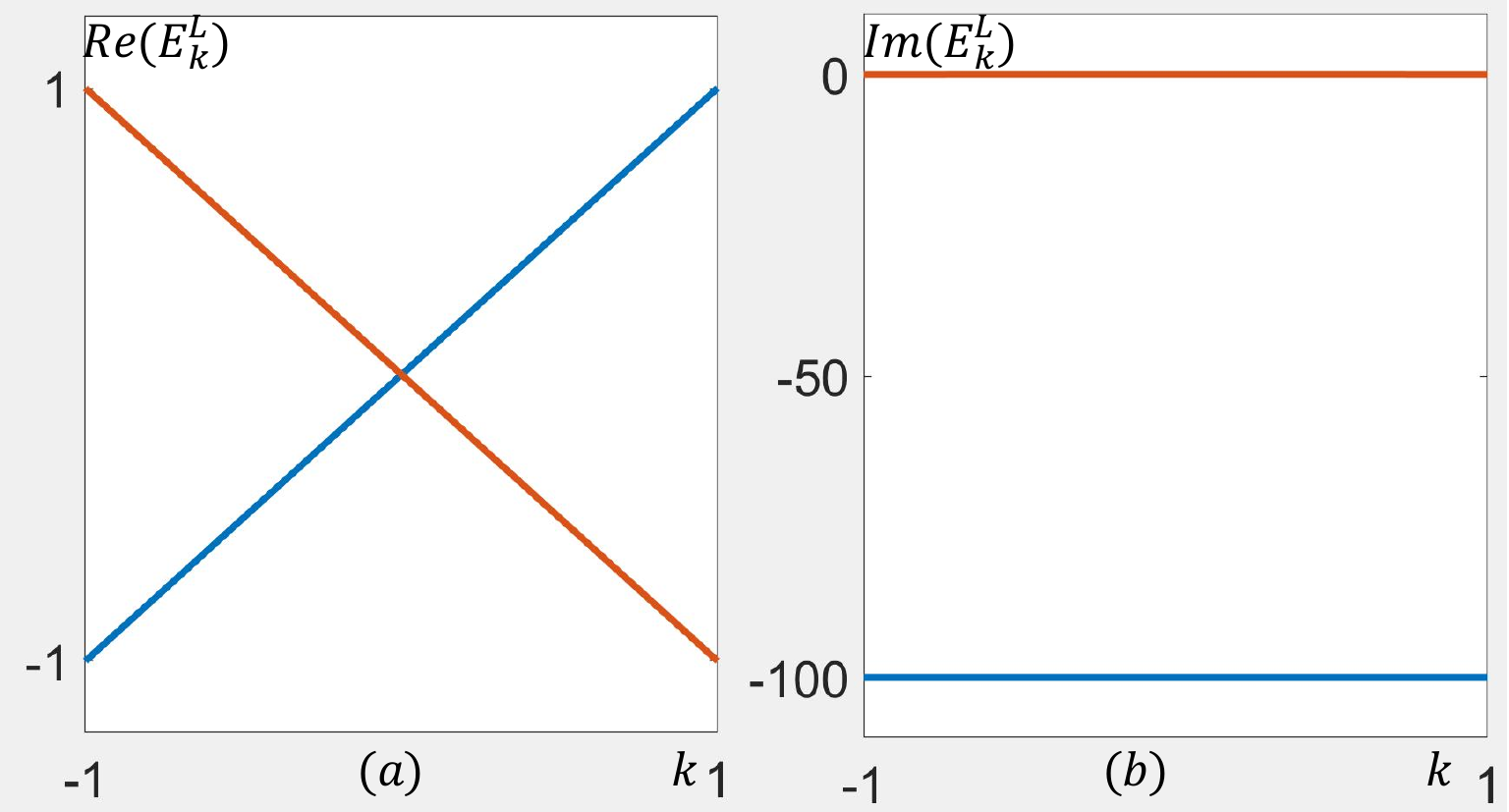}
\caption{\label{FigWH}
Illustration of band dispersion for the low energy effect Hamiltonian of chiral fermion in the oriented lattice. Red line: left-moving fermion; blue line: doubled fermion. We set $a=0.02$ and $V=1$.}
\end{figure}

The topological nature of the lattice-version index theorem suggests that the zero mode of our non-hermitian model is robust against local perturbations.  To check this, we consider a low energy effective Hamiltonian with coupling between the two kinds of fermions
\begin{align}
H=\sum_k &-(k+i\frac{k^2a}{2} ) c^\dagger_{L,0,k} c_{L,0,k} + (k - \frac{2i}{a}) c^\dagger_{L,\pi,k} c_{L,\pi,k} \nonumber\\& + V c^\dagger_{L,0,k} c_{L,\pi,k} + V c^\dagger_{L,\pi,k} c_{L,0,k}, \nonumber
\end{align}
where $c_{L,0,k}$ and $c_{L,\pi,k}$ corresponds to the left-moving fermion around $k = 0$ and the doubled fermion around $k = \pi/a$, respectively.
For small enough $a$ with $ak \ll 1, aV \ll 1$, we have
\begin{align*}
\epsilon_+ \approx -k-\frac{V^2+k^2}{2} a i,\quad \epsilon_- \approx k -\frac{2i}{a}.
\end{align*}
Thus, the perturbation does not open a gap.  It merely changes the imaginary part of the energy (see fig.\ref{FigWH} for an example), or the inverse lifetime of the fermions.  However, the qualitative behavior of lifetime of the fermions, i.e. the left-moving fermion has infinite lifetime in the continuum limit, while the lifetime of the right-moving fermion is zero in the continuum limit, are unchanged.

We also check the disorder effect by adding $\frac{i}{a}\sum_j \alpha_j c_{L,j}^\dagger c_{L,j} $ terms into the Hamiltonian \eqref{Hamiltonian} with randomly generated $\alpha_i \in [0,0.1]$. Our numerical result shows that such a perturbation does not open a gap either.  Thus, the zero mode of the oriented lattice model is much more robust than the one of the symmetric lattice model against the local perturbations.  

\section{Discussion and Conclusion} In summary, we have constructed a local non-hermitian 1D lattice model with a complex spectrum. We have demonstrated that this model describes a left-moving chiral fermion in various approaches, and shown that it is stable against local perturbations.   
Our results suggest that with the help of non-hermiticity, an anomalous field theory may be realized in a non-hermitian lattice model in the same spacetime dimension in the  continuum limit and low energy regime, where the non-hermiticity mimics a coupling between the lattice model and some kind of environments. In our example, the continuum limit means fixing $k$ and  $L=Na$, and letting $a\rightarrow 0, N\rightarrow \infty$ ; the low energy regime means $|kL|$ is of order 1 or less. \footnote{The low energy regime for doubled fermion should be $ka\rightarrow\pi$, then we can replace $k$ by $k+\pi/a$ to reset $k$ to be around 0.} Suggested by \eqref{SpectrumLimit},
the energy of doubled fermion has an infinite imaginary part and the mode damps out fast. While, the energy of left-moving fermion has a vanishing imaginary part, thus the mode decoupled from the environment represents a left-moving free fermion. That is how the low energy effective free fermion CFT emerges.

Finally, let us address the problem of possible experimental implementations of our non-hermitian lattice model. It is known that the non-hermitian lattice model may be realized with ultracold atoms in optical lattices. By taking the gauge field $A_1=\pi/a$ in eqn. \eqref{Hamiltonian}, this model matches exactly the eqn. F(4) in Ref. \cite{Gong2018} with $\kappa=-2J=1/a$, which has been proposed to be realizable in a system consists of two parallel fine-tuning optical lattices.  Alternatively, the model can also be simulated with electric circuits with diodes that induce left-right asymmetry and electrical inductors for the imaginary chemical potential 
 \cite{PhysRevB.99.201411,PhysRevApplied.13.014047}.
The zero modes can be detected via prominent two-point impedance peaks. 

 Our theory indicates that a chiral theory may be realized by coupling to some kind of environments instead of a one-dimensional higher bulk. This provides an alternative possibility, besides the spontaneous symmetry breaking, for the origin of at least some chiral phenomena in nature.

Notice added: After the completion of our work, we noticed a recent paper discussed similar issues about realizing anomalous theory on lattice model by Meng Cheng and Nathan Seiberg \cite{cheng2022lieb}.  However, the details are quite different. They mainly focus on 1-dimensional bosonic spin chain, while we work on 1-dimensional fermionic model.

\acknowledgments
We are grateful to the helpful discussions with Zheng-Cheng Gu and Ling-Yan (Janet) Hung.  This work was supported by National Key Research and Development Program of China (Grant No. 2016YFA0300300), NSFC (Grants No. 11861161001), the Science, Technology and Innovation Commission of Shenzhen Municipality (No. ZDSYS20190902092905285), and Center for Computational Science and Engineering at Southern University of Science and Technology.

\bibliography{bib}
\bibliographystyle{JHEP}
\end{document}